\documentclass[12pt]{article}
\usepackage{newtxtext,newtxmath}
\usepackage{graphicx}
\usepackage{braket}
\usepackage[letterpaper,margin=1in]{geometry}
\renewenvironment{abstract}
	{\quotation}
	{\endquotation}

\date{}

\makeatletter
\renewcommand{\fnum@figure}{\textbf{Figure\,\thefigure}}
\renewcommand{\fnum@table}{\textbf{Table \thetable}}

\usepackage{xcolor}

\usepackage[normalem]{ulem}

\usepackage{pdfpages}
\usepackage{pgffor}
\makeatletter 
\AtBeginDocument{\let\LS@rot\@undefined} 
\makeatother

\usepackage{scicite}
\usepackage{url}

\newtheorem{theorem}{Theorem}
\newtheorem{corollary}{Corollary}[theorem]

\newtheorem{lem}{Lemma}

\def\scititle{
    \fontsize{18pt}{20pt}\selectfont Engineered Randomness for Ubiquitous Quantum-Enhanced \\
    Metrology in Exponential-Dimensional Manifolds
}
\title{\bfseries \boldmath \scititle}
\author{
	Yaoming Chu$^{1,2}$,
	Baiyi Yu$^{3,4,1}$,
	Hartmut Häffner$^{3,4}$,\and
	Markus Heyl$^{5,6}$,
	Nathan Goldman$^{7,8,9}$,
	Jianming Cai$^{1,2,\ast}$\and
	\small$^{1}$ Center for Intelligence and Quantum Science (CIQS), \and
	\small International Joint Laboratory on Quantum Sensing and Quantum Metrology, School of Physics,\and
	\small Huazhong University of Science and Technology, Wuhan 430074, China.\and
	\small$^{2}$Hubei Key Laboratory of Gravitation and Quantum Physics, \and
	\small Institute for Quantum Science and Engineering, \and
	\small Huazhong University of Science and Technology, Wuhan 430074, China.\and
	\small$^{3}$Department of Physics, University of California, Berkeley, CA 94720, USA. \and
	\small$^{4}$Challenge Institute for Quantum Computation, University of California, Berkeley, CA 94720, USA. \and
    \small$^{5}$Theoretical Physics III, Center for Electronic Correlations and Magnetism, Institute of Physics, \and
	\small University of Augsburg, Universitätsstr. 12a, 86159 Augsburg, Germany. \and
    \small$^{6}$Centre for Advanced Analytics and Predictive Sciences (CAAPS),\and
	\small University of Augsburg, Universitätsstr. 12a, 86159 Augsburg, Germany. \and
	\small$^{7}$CENOLI, Universit\'e Libre de Bruxelles, CP 231, Campus Plaine, B-1050 Brussels, Belgium\and
    \small$^{8}$International Solvay Institutes, 1050 Brussels, Belgium\and
    \small$^{9}$Laboratoire Kastler Brossel, Coll\`ege de France, CNRS,\and  
    \small ENS-Universit\'e PSL, Sorbonne Universit\'e, 11 Place Marcelin Berthelot, 75005 Paris, France\and
	\small$^\ast$Corresponding author. Email: jianmingcai@hust.edu.cn
}

\begin{document} 
\maketitle
\newpage
\begin{abstract} \bfseries \boldmath
The exponential growth of many-body Hilbert space presents a fundamental barrier to quantum technology, obscuring the search for physically significant states within an astronomically vast landscape. Consequently, resources for quantum-enhanced metrology have been largely confined to the symmetric subspace whose dimensionality scales only polynomially with the particle number—leaving the vast majority of the Hilbert space largely unexplored and poorly understood. Here we challenge this paradigm by demonstrating that metrological advantage can arise as a ubiquitous feature across exponential-dimensional manifolds. By tailoring the first-moment structure of random unitaries, we uncover dense manifolds of engineered random states (ERSs) where Heisenberg-limited scaling emerges as a statistically generic property. This ubiquity endows these resource states with inherent resilience against parameter disorder. We experimentally validate this framework on a trapped-ion processor, achieving a metrological enhancement of $6.98 \pm 0.38$ dB beyond the standard quantum limit. Potential applications extend to diverse platforms, ranging from superconducting circuits and waveguide QED to solid-state spins and polar molecules. Our results establish a powerful paradigm where quantum-enhanced precision can be harvested from the exponential vastness of the Hilbert space.
\end{abstract}

\noindent
The dimension of a quantum system's Hilbert space grows exponentially with the number of particles $N$, quickly becoming astronomically vast \cite{Feynman1982}: for $N=300$ qubits, the $2^{300}$ available basis states already exceed the estimated number of atoms in the observable universe. This exponential complexity provides the foundation for many-body physics and quantum-enhanced technologies, yet simultaneously defines the central challenge for navigating and exploring this space to find states with useful properties. At a more fundamental level, uncovering the underlying structure of the Hilbert space—its geometry \cite{Nielsen2006,Brown2023} and governing constraints \cite{Lieb1972,Eisert2010,Lami2023}—and identifying its physically significant manifolds stand as a critical frontier.

Historically, the search for metrologically useful states has largely centered on the symmetric subspace \cite{Degen2017,Pezze2018}, yielding canonical resources such as Dicke \cite{Dicke1954}, Greenberger-Horne-Zeilinger (GHZ) \cite{GHZ1989} and squeezed states \cite{Kitagawa1993}. Although quantum advantage is identified within this narrow niche of polynomial dimensionality \cite{Oszmaniec2016}, generic many-body states in the exponentially vast Hilbert space are widely believed to offer little metrological usefulness. Known exceptions—for example, ground states of specific local Hamiltonians \cite{Frerot2018,Niezgoda2021,Comparin2022} or states generated through fine-tuned quantum evolution \cite{Bornet2023,Eckner2023,Franke2023,Block2024,Wu2025,Gao2025}—are rare and exist only as isolated singularities. These facts conceptually raise a pivotal question: {\it Does the Hilbert space embody structured manifolds where quantum advantage ubiquitously persists across much higher—potentially even exponential—dimensions}?

\begin{figure}[!t]
	\centering
	\includegraphics[width=125mm]{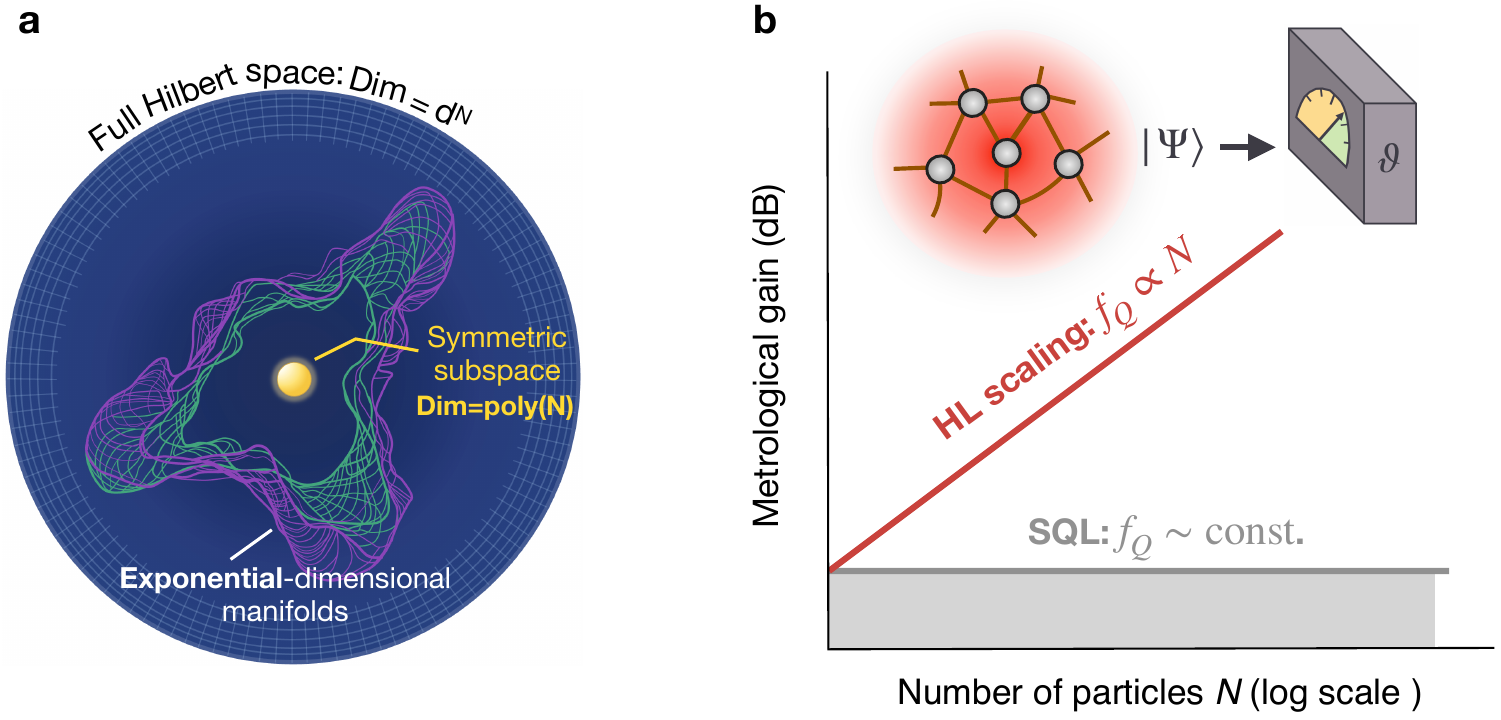}
      \caption{{\bf Quantum-enhanced metrology from the exponentially vast Hilbert space.} ({\bf a}) Generic Haar-random states offer negligible metrological usefulness, thus it is critical to identify the structured manifolds embedded in the Hilbert space that yield quantum advantage. ({\bf b}) The metrological usefulness of an input state $|\Psi\rangle$ for estimating a parameter $\vartheta$ is quantified by the QFI density, $f_Q = F_Q/N$, which determines the fundamental precision floor via the quantum Cramér-Rao bound, $\delta \vartheta \geq (N f_Q)^{-1/2}$. The constant ($f_Q\sim\mathrm{const.}$) and linear growth ($f_Q\propto N$) behaviors correspond to the SQL and Heisenberg-limited (HL) scalings, respectively. }
    \label{fig:model}
\end{figure}

Here, we show that physically significant manifolds, which enables quantum-enhanced metrology to surpass the standard quantum limit (SQL), can be identified by leveraging cleverly engineered randomness across the Hilbert space. Specifically, by shaping the first moments of unitary ensembles beyond the trivial Haar-randomness \cite{Oszmaniec2016,Guhr1998}, we theoretically find that Heisenberg-limited (HL) scaling—representing a fundamental $N^{1/2}$ precision enhancement over the SQL—unexpectedly emerges as a statistically generic property of exponential-dimensional manifolds (see Figure\,\ref{fig:model}a). Our findings affirmatively answer the posed question that quantum advantage is not a rare, symmetry-protected exception, but a ubiquitous feature that permeates the exponential reaches of the Hilbert space via engineered randomness.

We experimentally validate quantum-enhanced metrology on a trapped-ion processor, achieving a phase measurement gain of 6.98(±0.38) dB beyond the SQL using 10-qubit engineered random states (ERSs). Crucially, we find that metrological advantage is a universal attribute of the ERS manifold, granting the strategy remarkable robustness against parameter disorder that otherwise severely degrades conventional protocols for resource state generation. These results not only substantially broaden the scope of quantum metrology, but also illuminate our understanding of the fundamental geometry of entanglement and the structural landscape of many-body quantum states. Our work thus opens an avenue toward efficient harvesting of useful states from the exponentially vast Hilbert space through properly engineered randomness.

\section*{Main result}
To set the stage, we investigate a generic quantum system composed of $N$ particles, each equipped with a local Hilbert space of dimension $d$. The power of quantum metrology is fundamentally dictated by how distinguishable a quantum state becomes under the continuous encoding of a parameter $\vartheta$. This distinguishability is rigorously quantified by the quantum Fisher information (QFI), $F_Q [\rho; \mathcal{O}]$, a metric that defines the distance between an input state $\rho$ and its parameter-translated counterpart $e^{-i \vartheta \mathcal{O}} \rho e^{i \vartheta \mathcal{O}}$ \cite{Braunstein1994, Pezze2018}. Here, $\mathcal{O}$ is the Hermitian generator associated with the translation of $\vartheta$; without loss of generality, we adopt a local generator: $\mathcal{O} = \sum_{m=1}^N \mathcal{O}_m$, where the operator $\mathcal{O}_m$ acting on the $m$-th particle has a spectrum of unit width. For a pure state $\rho=|\Psi\rangle\langle \Psi|$, the QFI is determined by its susceptibility to the generator $\mathcal{O}$, yielding the elegantly simple form as follows
\begin{equation}
\label{Eq:QFI}
F_Q (|\Psi\rangle; \mathcal{O}) = 4 \left(\langle \Psi |\mathcal{O}^2 |\Psi\rangle -\langle \Psi |\mathcal{O} |\Psi\rangle^2\right).
\end{equation}
The operational significance of the QFI lies in its direct relation to the ultimate metrological performance: it establishes a fundamental limit on the achievable precision, $\delta \vartheta$, via the so-called quantum Cramér-Rao bound, $(\delta \vartheta)^{-2}\leq F_Q [|\Psi\rangle; \mathcal{O}]$ \cite{Braunstein1994}. Crucially, the QFI also connects metrological precision directly to the depth of multipartite entanglement. A QFI density $f_Q = F_Q/N$ exceeding an integer $k$ verifies that the system possesses at least ($k+1$)-partite entanglement \cite{Hauke2016,Zanardi2006,Garttner2018,Smith2016,Desaules2022,Chu2024}. Consequently, achieving HL precision ($F_Q \propto N^2$) strictly requires genuine multipartite entanglement, characterized by an extensive QFI density$f_Q \propto N$  (see Figure\,\ref{fig:model}b).

Below, we demonstrate—through two key findings—that many-body states generated via properly engineered randomness can generically unlock the HL precision scaling from the exponentially vast Hilbert space. Specifically, our framework generates metrological states by evolving an initial state $|\Psi\rangle$ under a unitary transformation $U$. Without loss of generality, hereafter $|\Psi\rangle$ is assumed to be a $N$-particle product state: $|\Psi\rangle=\otimes_{m=1}^N \ket{\psi}_m$, where each component is the eigenstate of the local traceless Hermitian generator $\mathcal{O}_m$ with a positive eigenvalue $\lambda$, i.e. $\mathcal{O}_m \ket{\psi}_m = \lambda \ket{\psi}_m$.
Our first finding is a class of {\it $\alpha$-random unitary ensembles}—denoted by $\mathcal{U}_\alpha$—that are distinguished by the characteristic first-moment structure
\begin{equation}
\label{Eq:First-Moment}
\mathbb{E}_{U\sim \mathcal{U}_\alpha} [U_{i_1 j_1} U^{*}_{i_2,j_2}] =\alpha \delta_{i_1 j_1} \delta_{i_2,j_2} + (1-\alpha) \Delta_{i_1,i_2,j_1,j_2}.
\end{equation}
Here, $\alpha\in (0,1)$ is the defining parameter of the ensemble, and the second term corresponds to a completely depolarizing channel for a quantum operator $\mathcal{A}$,
\begin{equation}
\begin{aligned}
\mathcal{D}(\mathcal{A})\equiv \sum_{i_1 i_2 j_1 j_2} \Delta_{i_1,i_2,j_1,j_2} \mathcal{A}_{i_2 i_1} |j_2\rangle \langle j_1| = c \frac{\mathbb{I}}{D},
\end{aligned}
\end{equation}
where the constant $c=\mathrm{tr}(\mathcal{A})$ is fixed by invoking the unitarity of $U$ \cite{supplement}, and $D=d^N$ is the Hilbert space dimension. Crucially, $\alpha$-random unitary ensembles have a unique feature, as stated in the following lemma. 

\begin{lem}
\label{lem}
An $\alpha$-random unitary ensemble $\mathcal{U}_\alpha$ contracts a quantum operator $\mathcal{A}$ by a factor of $\alpha$ in an average sense
\begin{equation}
\mathbb{E}_{U\sim \mathcal{U}_\alpha} [U^\dagger \mathcal{A} U] = \alpha \mathcal{A} +(1-\alpha) \mathrm{tr}(\mathcal{A}) \frac{\mathbb{I}}{D}.
\end{equation}
\end{lem}

In contrast to the standard Haar-randomness leading to $\alpha=0$, our engineered randomness breaks this triviality by introducing the first $\alpha$-contraction term  [with $\alpha\in(0,1)$] for arbitrary quantum operators. This non-trivial universal contraction is the key mechanism essentially underlying the emergence of a HL precision scaling achieved by the generated ERSs. As an illustrative example, we consider a $N$-qubit system initially prepared in a fully polarized state along the $z$ axis (i.e. $|\Psi\rangle=\ket{0}^{\otimes N}$) and take the local Hermitian generator associated with the parameter change as $\mathcal{O}_m=\sigma_z^{(m)}/2$, where $\sigma_\mu^{(m)}$ with $\mu = x, y, z$ are Pauli operators supported on the $m$-th qubit. Applying random unitaries from $\mathcal{U}_\alpha$ contracts the expectation values of $J_z= \sum_{m=1}^N \sigma_z^{(m)}/2$ and $J_z^2$ by a factor of $\alpha$ based on Lemma 1. Therefore, the two terms in Eq.\,\eqref{Eq:QFI} for $\mathcal{O}=J_z$ are reduced by $\alpha$ and $\alpha^2$, yielding an average QFI of $\alpha (1-\alpha) N^2+O(N)$ that reaches the HL scaling. 

The result is rigorously formalized in the following theorem, with proofs outlined in the Appendix and detailed in the Supplementary Materials \cite{supplement}.

\begin{theorem}
Let $|\Psi_U\rangle = U|\Psi\rangle$ be the family of ERSs generated by $U$ from an $\alpha$-random unitary ensemble $\mathcal{U}_\alpha$. Then, the quantum Fisher information of these states associated with the generator $\mathcal{O} = \sum_{m=1}^N \mathcal{O}_m$ attains the following ensemble average,
\begin{equation}
\label{Eq:QFImean}
\mathbb{E}_{U\sim \mathcal{U}_\alpha} [F_Q(|\Psi_U\rangle;\mathcal{O})] = 4 \alpha (1-\alpha) \lambda^2 N^2 + \Theta(N),
\end{equation}
provided that the variance of $f_U=\langle \Psi_U|\mathcal{O} |\Psi_U\rangle$ scales at most as $\Theta(N)$, namely $\mathbb{E}_{U\sim \mathcal{U}_\alpha}[f_U^2]-\mathbb{E}^2_{U\sim \mathcal{U}_\alpha}[f_U]\leq \Theta(N)$. Here, $\Theta(\xi)$ indicates an asymptotic scaling identical to $\xi$ for large $\xi$.
\label{Theorem:1}
\end{theorem}

Importantly, Theorem \ref{Theorem:1} establishes a principled and systematic route toward ERS manifolds with superior average metrological usefulness. We emphasize that these ERSs are largely disjoint from the symmetric subspace \cite{supplement}, thus representing fundamentally new types of metrologically useful states.
Building upon this framework, we next turn to present our second key finding, which addresses the distribution of ERS performance by revealing a striking concentration-of-measure phenomenon in high-dimensions: the QFI of most ERSs generated by the $\alpha$-random unitary ensemble can cluster within a thin shell around their average.

\begin{theorem}
    Under the same setting of Theorem \ref{Theorem:1}, the quantum Fisher information of ERSs generated by an $\alpha$-random unitary ensemble concentrates to the Heisenberg-limited scaling,
    \begin{equation}
    \mathrm{Pr}\left(F_Q(|\Psi_U\rangle; \mathcal{O}) <\Theta(N^2) \right) \leq \exp\left(-\Theta\left(\frac{D}{\mathrm{poly}(N)}\right)\right),
    \end{equation}
    provided that the expectation value function $f(U;\mathcal{A}) = \langle \Psi_U|\mathcal{A} |\Psi_U\rangle$ for $\mathcal{A}\in \{\mathcal{O},\mathcal{O}^2\}$ is $L_\mathcal{A}$-Lipschitz continuous and satisfies the following inequality for $\forall \epsilon > 0$, 
    \begin{equation}
    \label{Eq:EVconcentration}
    \mathrm{Pr}\Big(\left|f(U;\mathcal{A})- \mathbb{E}_{U\sim \mathcal{U}_\alpha} [f(U;\mathcal{A})]\right| \geq \epsilon \Big) \leq 2\exp\left(-\frac{c\epsilon^2}{L_\mathcal{A}^2}\right),
    \end{equation}
    where the probability $\mathrm{Pr}(\cdot)$ is taken over $\mathcal{U}_\alpha$ and $c$ represents a universal positive constant with the ratio $c/L_\mathcal{A}^2=\Theta(D/\mathrm{poly}(N))$. Here, the symbol $\mathrm{poly}(\cdot)$ denotes an arbitrary polynomial.
    \label{Theorem:2}
\end{theorem}

We remark that the condition in Eq.\,\eqref{Eq:EVconcentration} reflects a generic and fundamental feature of high-dimensional ensembles, which can be rigorously established for diverse measures with positive Ricci curvature \cite{Ledoux2001,Anderson2009,Bakry2014}. 
Theorem \ref{Theorem:2} demonstrates that the ERSs generated by $\alpha$-random unitary ensembles constitute a physically significant manifold within the Hilbert space for quantum-enhanced metrology, and the probability of encountering a ERS that cannot achieve the HL scaling is exponentially vanishing. Together with Theorem \ref{Theorem:1}, it leads to our main result that {\it many-body ERSs are capable of generically achieving enhanced precision at the Heisenberg-limited scaling}. This suggests that metrologically useful states are not rare, fine-tuned anomalies; rather, they can emerge as typical outcomes when sampling from certain manifolds in the Hilbert space.

\begin{figure}[!t]
	\centering
	\includegraphics[width=136mm]{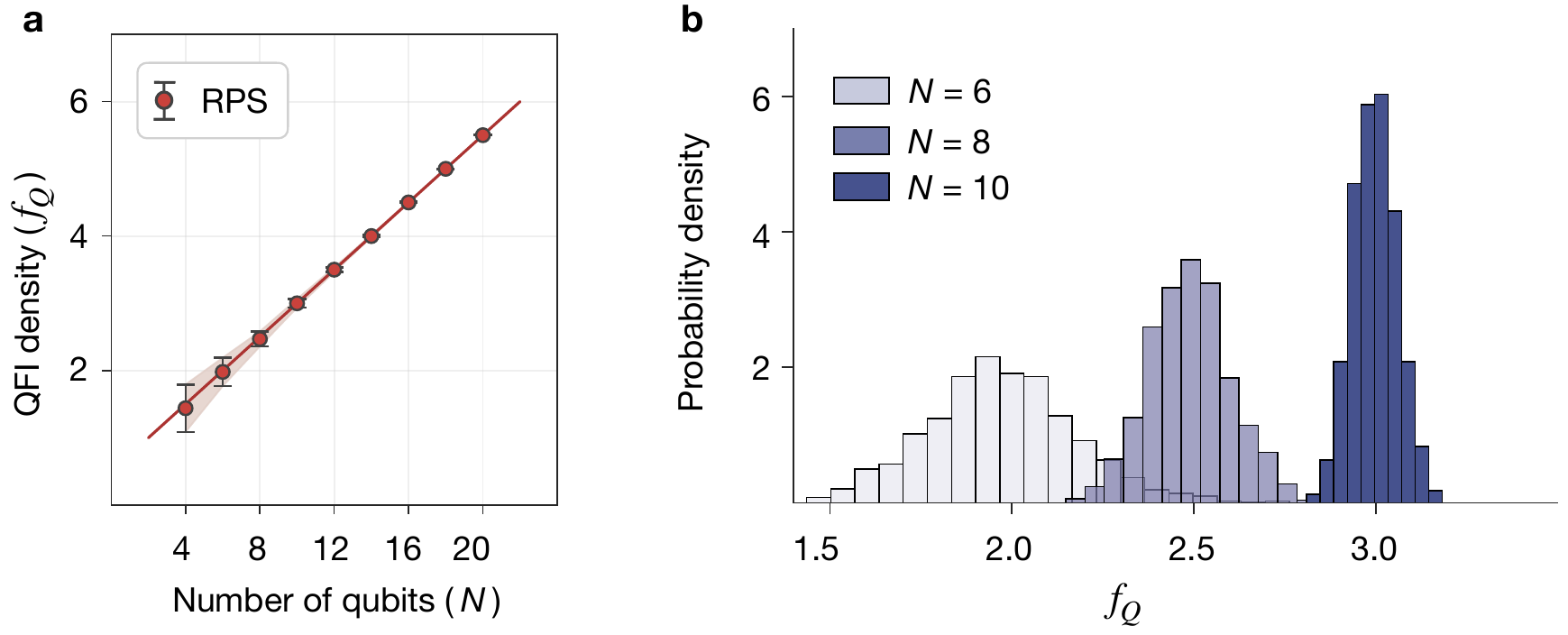}
      \caption{{\bf Heisenberg-limited scaling emerging from engineered random unitary ensembles.} ({\bf a}) At $\alpha=1/2$, the average QFI densities for RPSs—a specific example of ERSs—exhibit excellent agreement with the theoretical prediction of HL scaling $\mathbb{E}[f_Q] = N/4 + 1/2$ (solid line). Each data point represents an average over 200 exact random numerical realizations. The shaded regions denote the statistical spread, which rapidly diminishes with system size, revealing the concentration-of-measure phenomenon.  ({\bf b}) Probability density distributions from 2000 realizations at different system sizes further highlight the concentration of $f_Q$ in high dimensions. }
    \label{fig:RPS}
\end{figure}

\section*{Examples of ERS manifolds}
Theorem 1 and 2 provide a guiding framework for finding Heisenberg-limited ERS manifolds in the exponentially vast Hilbert space. To demonstrate the feasibility of $\alpha$-random unitary ensembles to generate metrologically useful ERSs, we treat the general form of a unitary operator $U$ via a similarity transformation,
\begin{equation}
\label{Eq:Construction}
U = V \Phi V^\dagger,
\end{equation}
where $V$ and $\Phi$ denote the transformation and core unitaries, respectively. We next present two explicit examples of $\alpha$-random unitary ensembles that satisfy the prescribed first-moment statistical identities in Eq.\,\eqref{Eq:First-Moment}.

\subsubsection*{Random phase state}
The first example is a simple and interesting class of metrologically useful states of a $N$-qubit system, namely the random phase states (RPSs), written as 
\begin{equation}
\label{Eq:RPS}
|\Psi (\varphi_1,\cdots,\varphi_D)\rangle = \frac{1}{\sqrt{D}}\sum_{b=1}^D e^{i \varphi_b} |b\rangle,
\end{equation}
where $\{|b\rangle\}_{b=1}^D$ spans the computational basis and $\{\varphi_b\}_{b=1}^D$ is a set of independent random phases. The RPSs are generally believed to offer no quantum advantage. Surprisingly, we will show that they can in fact serve as a potent resource for achieving the HL precision scaling.  

Starting from an initial product state aligned along the $x$-axis, namely $|\Psi\rangle=\ket{\boldsymbol{x}}=\ket{+}^{\otimes N}$, RPSs can be generated via a diagonal $\alpha$-random unitary ensemble, $\Phi = \mathrm{diag}[\{e^{i \varphi_b}\}_{b=1}^D]$, where each random phase is drawn independently from a distribution $p(\varphi_b)$. The corresponding ensemble contraction factor [see Eq.\,\eqref{Eq:First-Moment}], is given by \cite{supplement}
\begin{equation}
\label{Eq:alpha}
\sqrt{\alpha} = \mathbb{E}_{\varphi_b\sim p(\varphi_b)}[\exp(i\varphi_b)].
\end{equation}
Theorem 1 guarantees that the ensemble-averaged QFI exhibits HL scaling as long as the factor $\alpha$ lies within $(0,1)$. Rather than adopting the conventional unitary ensemble \cite{Nakata2013} with each phase uniformly distributed within $[0,2\pi]$ that trivially results in $\alpha=0$, we consider a general normal distribution, for which $\alpha=\exp(-\sigma^2)$ with $\sigma$ the standard deviation of $\varphi_b$.

We further establish the typicality of HL scaling by situating this RPS class within the geometry of high-dimensional Gaussian measure spaces. Identifying the collective spin operator $J_x$ as the generator of parameter change, we explicitly evaluate the concentration behavior in Eq.\,\eqref{Eq:EVconcentration} for $\mathcal{A}\in\{J_x, J_x^2\}$. It can be shown that $c\propto \sigma^{-2}$, while the Lipschitz constants are upper bounded by $L_\mathcal{A}\leq N^\gamma/D^{1/2}$ with $\gamma=1$ and $2$, respectively \cite{supplement}. Consequently, Theorem \ref{Theorem:2} ensures the tight concentration of the QFI at its mean, establishing the metrological advantage as a ubiquitous property of the entire RPS manifold (see Figure\,\ref{fig:RPS}), as stated by the following corollary.

\begin{corollary}
For any $\alpha \in (0,1)$, nearly all random phase states can achieve quantum-enhanced precision at the Heisenberg-limited scaling. 
\label{theorem:rps}
\end{corollary}

Remarkably, we can also rigorously prove that the RPS manifold necessitates an exponential embedding dimension of $2^N$; namely, it cannot be confined to any Hilbert subspace of dimension smaller than $2^N$ \cite{supplement,Mityagin2020}. This fact signifies that the RPS manifold exploits the full exponential complexity of the Hilbert space—a stark departure from the polynomial constraints of the conventional symmetric subspace.

\subsubsection*{Chimera-superposition states}
We apply our findings to the second example with the transformation unitaries $V$ in Eq.\,\eqref{Eq:Construction} chosen to be Haar-random on the special unitary group $\mathrm{SU}(D)$ \cite{Guhr1998}. Interestingly, the class of states generated from this Haar-random construction admit the following decomposition,
\begin{equation}
\label{Eq:ChimeraState}
|\Psi_U\rangle = \sqrt{\alpha} e^{i\beta} |\Psi\rangle + \sqrt{1-\alpha} |\Psi_{\mathrm{H}}^{(1)}\rangle,
\end{equation}
with $\beta=\mathrm{arg}[\mathrm{tr}(\Phi)]$, and the contraction factor $\alpha $ [see Eq.\,\eqref{Eq:First-Moment}] is given by 
\begin{equation}
\alpha =\frac{1}{D} |\mathrm{tr}(\Phi)|^2.
\label{Eq:CSS_alpha}
\end{equation} 
The first component $|\Psi\rangle$ represents the initial product state. The second component, $|\Psi_{\mathrm{H}}^{(1)}\rangle$, exhibits first-order Haar-randomness, characterized by vanishing polarization along any arbitrary directions. This component can be further generalized to be exact Haar-random states \cite{supplement}. The resulting states therefore embody a new form of quantum superposition, intertwining an ordered contribution (such as fully polarized states) with a highly-random part. We thus designate these configurations as {\it Chimera-superposition states} (CSSs). 

The two constituents of a CSS differ markedly in their physical character---for instance, in state complexity \cite{Aaronson2020,Brown2023}---and cannot be connected through simple observables such as $\mathcal{A} = J_z$ or $\mathcal{A} = J_z^2$, namely $|\langle \Psi|\mathcal{A}|\Psi_{\mathrm{H}}^{(1)}\rangle|^2 = \Theta(1/D)$ \cite{supplement}. As a consequence, expectation values of such observables can be approximated as
\begin{equation}
\langle \Psi_U|\mathcal{A}|\Psi_U\rangle \approx \alpha \langle \Psi|\mathcal{A}|\Psi\rangle + (1-\alpha) \langle \Psi_{\mathrm{H}}^{(1)}|\mathcal{A}|\Psi_{\mathrm{H}}^{(1)}\rangle,
\end{equation} 
which is naturally consistent with the operator contraction in Lemma 1, since the second term on the right-hand side is typically negligible. 
Building on Theorem \ref{Theorem:1} and \ref{Theorem:2}, we can similarly extend the concentration of superior metrological usefulness 
to this CSS manifold by identifying $c=D/4$ and $L_{\mathcal{A}}\propto \mathrm{poly}(N)$ for $\mathcal{A}\in \{J_z, J_z^2\}$ [see Eq.\,\eqref{Eq:EVconcentration}] under the Haar measure on $\mathrm{SU}(D)$.

\begin{corollary}
For any $\alpha \in (0,1)$, nearly all Chimera-superposition states can achieve quantum-enhanced precision at the Heisenberg-limited scaling.
\label{Theorem:SCRS}
\end{corollary}

This corollary provides a solid mathematical foundation for the ubiquitous quantum advantage inherent to the CSS manifold. It rigorously establishes that the metrological usefulness often attributed to complex scrambling dynamics \cite{Kobrin2024,Ge2025,Hu2026} is, in fact, a generic feature of engineered randomness. Similar results can be extended to the Gaussian unitary ensembles \cite{supplement,Erdos2014,Cotler2017}. Notably, similar to the RPSs, this CSS manifold also possesses an embedding dimension of $D=d^N$, matching the full exponential complexity of the Hilbert space \cite{supplement}. 

\begin{figure}[!t]
    \centering
    \includegraphics[width=160mm]{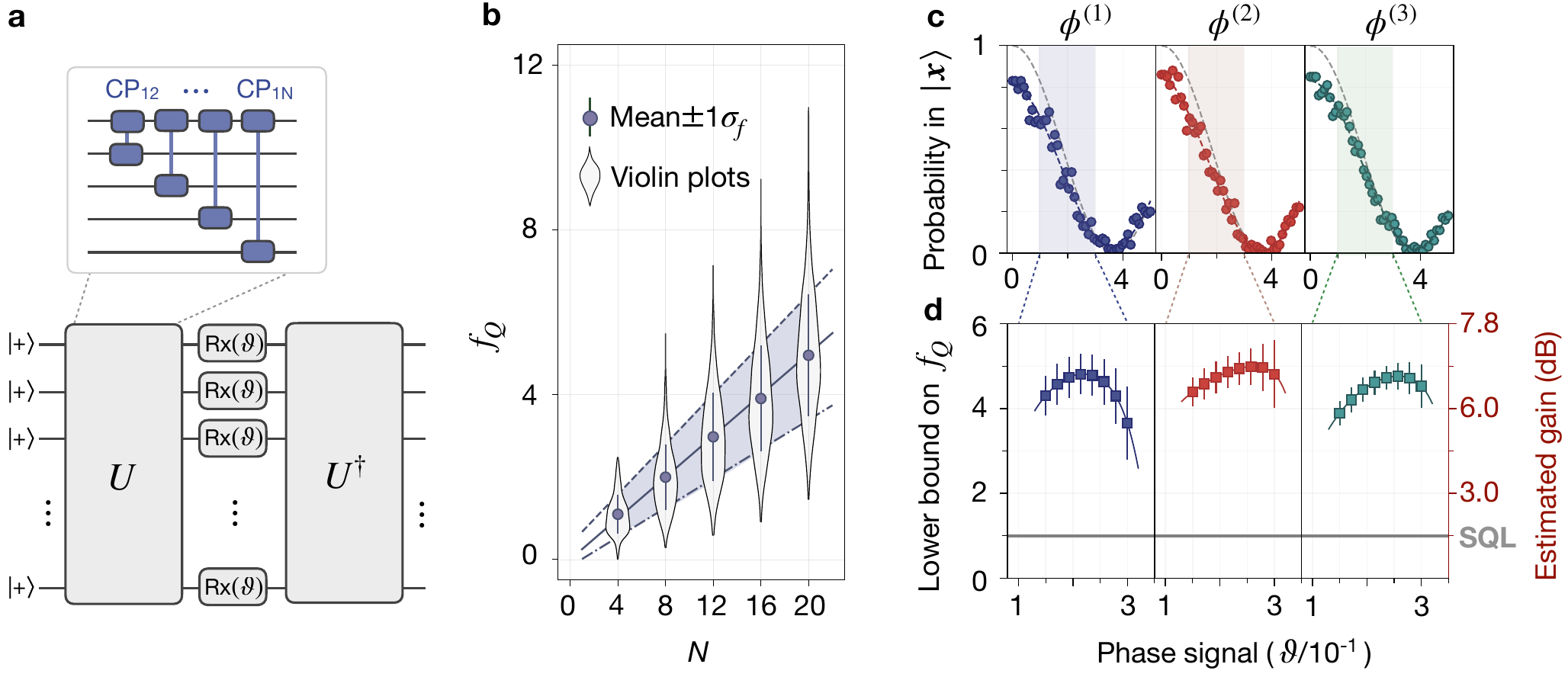}
    \caption{{\bf Quantum-enhanced metrology using ERSs in a trapped-ion platform.} {\bf (a)} Time-reversal sensing protocol with ERSs created by random quantum circuits (RQCs) consisting of two-qubit CP gates (see Appendix). {\bf (b)} Violin plots of $f_Q$ for varying system sizes, each based on 2000 random numerical realizations of the ERS. Mean values (filled circles) agree well with $\mathbb{E}[f_Q]=N/4$ (solid line), while $\mathbb{E}[f_Q]\pm\sigma_f$ also exhibit linear scaling with slopes 0.30 (dashed) and 0.18 (dash-dotted). {\bf (c)} Three 10-qubit RQCs, with different random interaction weights [see Eq.\,\eqref{Eq:RHE} and Appendix], are executed on IonQ's Forte-1 processor. After phase encoding and time reversal, the return probability $P(\vartheta)$ is recorded as a function of $\vartheta$. {\bf (d)} The lower bound on $f_Q$ of the experimentally generated ERS are inferred from the measured phase uncertainty. The corresponding metrological gain beyond the SQL is indicated, showing the quantum enhancement achievable with these states. In all panels, a standard deviation of $\sigma_\phi=0.5\pi$ is set for the interaction weight parameters.}
    \label{fig:CRPS}
\end{figure}

\section*{Experimental verification of enhanced metrology using ERSs}
Guided by the generic usefulness established for the RPS manifold, we extend our analysis to an easily accessible class of ERSs within $N$-qubit experimental platforms. Specifically, we consider a random Hamiltonian ensemble restricted to at most two-body Ising interactions,
 \begin{equation}
 \label{Eq:RHE}
 H = \sum_{(ij)\in e^*} \phi_{ij} \Big(\sigma_z^{(i)} \sigma_z^{(j)}- \sigma_z^{(i)}-\sigma_z^{(j)}\Big),
 \end{equation}
where $e^*$ denotes the edge set of  a star configuration. Each interaction weight $\phi_{ij}$ is independently sampled from a normal distribution with zero mean and a standard deviation such that $\mathbb{E}[\exp(4i\phi_{ij})]\ll 1$. These weights are mapped to the phases $\{\varphi_b\}$ in Eq.\,\eqref{Eq:RPS} via a $D\times (N-1)$ linear transformation matrix \cite{supplement}. Start from an initial state $\ket{\boldsymbol{x}}=\ket{+}^{\otimes N}$, we apply the unitaries $U= \exp(i H)$ via the random quantum circuits (RQCs) in Figure\,\ref{fig:CRPS}a. This procedure generates random weighted graph states \cite{Hartmann2007}—which essentially constitute a specific realization of the broader RPS class characterized by correlated random phases \cite{supplement}. We show in what follows that these ERSs also possess a significant metrological potential that has surprisingly remained hitherto unexplored. It is worth to emphasize that the mutual commutativity of all terms in Eq.\,\eqref{Eq:RHE} in principle allows for the parallel execution of all two-qubit entangling operations. Thus, this offers a scalable route toward the rapid preparation of large-scale ERSs achieving Heisenberg-limited quantum measurement precision \cite{Chu2023}.

Numerical simulations spanning a large number of such RQCs, presented in Figure\,\ref{fig:CRPS}b, reveal that a predominant fraction of the resulting states displays genuine metrological advantage: their QFI densities significantly exceed the SQL (i.e. $f_Q=1$) and the ensemble average follows the theoretical scaling of $\mathbb{E}[f_Q]=N/4$ \cite{supplement}. Crucially, while these circuit-generated ERSs differ from the exact RPS class due to correlations among random phases, the $\pm 1$ standard deviation bounds still faithfully maintain the HL scaling \cite{supplement}. This statistical spread also enables the efficient identification of a significant subset of ERSs with QFI densities well beyond the ensemble average.

In our experiments, we create three instances of such 10-qubit ERSs with distinct sets of random interaction weights on an IonQ trapped-ion processor, and demonstrate quantum-enhanced metrology by exploiting a time-reversal protocol \cite{Colombo2022,Liu2022}. In this protocol, the prepared ERS undergoes a collective phase shift $e^{i\vartheta J_x}$, where $\vartheta$ represents the parameter to be estimated, modeling the signal accumulated from an external field coupled to the probe in a standard sensing scenario. Following this interrogation stage, a time-reversed circuit $U^\dagger$ is applied to unravel the engineered correlations and convert the accumulated phase into a measurable population signal. The measured probabilities $P(\vartheta)$, of which the system returns to $|\boldsymbol{x}\rangle$, are presented in Figure\,\ref{fig:CRPS}c. From the regions of $P(\vartheta)$ with large slopes versus $\vartheta$, we extract a metrological gain of $6.98(\pm 0.38)$ dB beyond the SQL, giving rise to apparent quantum enhancement, as detailed in Figure\,\ref{fig:CRPS}d. We note that the realized ERSs are mixed states due to inevitable decoherence within the real quantum device. Their QFI exhibits a highly nonlinear and intricate dependence on the underlying density matrix \cite{Hauke2016, supplement}, rendering the direct evaluation intractable. In this situation, the observed measurement precision therefore provides a certified lower bound on the QFI density of these ERSs, as dictated by the quantum Cram\'er-Rao bound—which holds universally for both pure and mixed states \cite{Braunstein1994}.

\begin{figure}[!t]
    \centering
    \includegraphics[width=153mm]{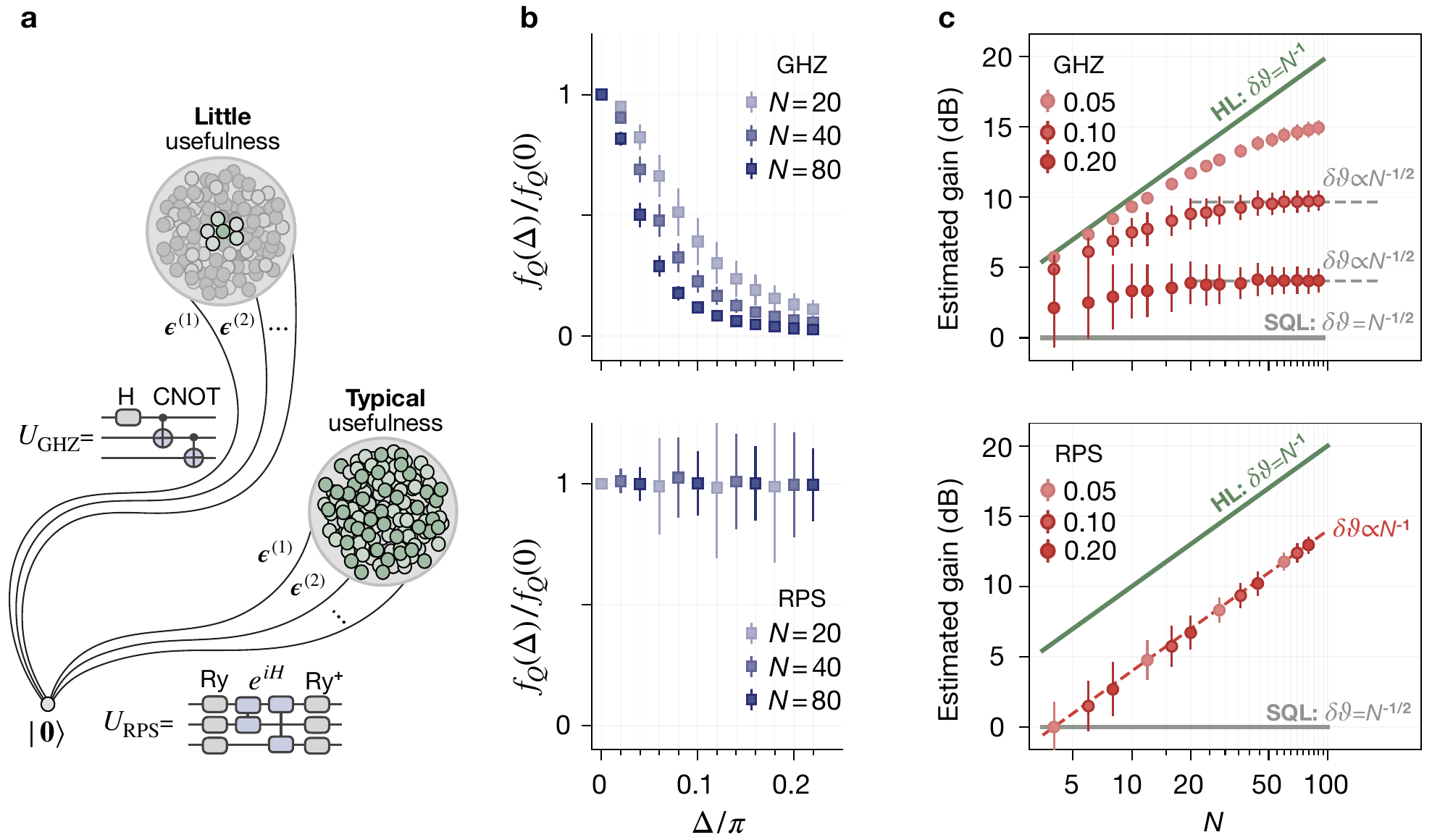}
    \caption{{\bf Numerical analysis of metrological resilience against parameter disorder.}  {\bf (a)} Schematic of parameter disorder (denoted by $\boldsymbol{\epsilon}$) introduced into the two-qubit CNOT gates for GHZ-state preparation and the Hamiltonian interaction weights [see Eq.\,\eqref{Eq:RHE}] for ERS generation. Each random instance is independently sampled from a normal distribution $\mathcal{N}(0, \Delta^2)$. {\bf (b)} Normalized QFI density (relative to the ideal case of $\boldsymbol{\epsilon}=0$) as a function of the disorder strength $\Delta$. The traditional protocol for the GHZ state generation exhibits a marked deterioration, whereas the performance of the ERS framework remains essentially invariant, demonstrating near-complete resilience to parameter disorder. {\bf (c)} Scaling of the metrological gain (estimated from the QFI) with the system size $N$. Symbols and curves from light to dark shades represent $\Delta/\pi=0.05, 0.1, 0.2$, with upper and lower panels corresponding to the GHZ and ERS scenarios respectively. For a fixed $\Delta$, the extracted gain of the perturbed GHZ states ceases to grow at large $N$, eventually falling back to the SQL yielding a precision of $\delta \theta\propto N^{-1/2}$. In stark contrast, the generated ERSs maintain the characteristic HL scaling ($\delta \theta = 2N^{-1}$) even under substantial disorder.}
    \label{fig:resilience}
\end{figure}

\section*{Metrological resilience against parameter disorder}
In contrast to conventional protocols, where metrological usefulness hinges on a singular configuration, the ERS generation affords substantial experimental flexibility by exploiting the typicality of HL scaling over a dense region of Hilbert space. Numerical results of Figure\,\ref{fig:resilience} illustrate this metrological resilience by examining the effect of parameter disorder on the ERS framework and the canonical GHZ-state preparation protocol. As examples, we consider the $\mathrm{CNOT}$ gates used for the GHZ state varied as $\mathrm{CNOT}^{(\mathrm{\epsilon})} = \exp[-i(\pi/4+\epsilon_i) \sigma_z^{(i)} \sigma_x^{(i+1)}]\exp[i \pi (\sigma_z^{(i)}+\sigma_x^{(i+1)})/4]$ \cite{Chow2011IBM,Sheldon2016IBM}, while the interaction weights in Eq.\,\eqref{Eq:RHE} for ERS creation are perturbed according to $\phi_{ij}^{(\epsilon)} = \phi_{ij} + \epsilon_{ij}$, with the baseline parameters $\{\phi_{ij}\}$ yielding a QFI density near the ensemble average to characterize the typical feature of the ERSs. Here, all perturbation parameters $\epsilon_i$ and $\epsilon_{ij}$ are independently drawn from a normal distribution with zero mean and standard deviation $\Delta$. With increasing $\Delta$, the estimated metrological gain of the perturbed GHZ states is severely compromised, causing the scaling to revert from the HL toward the SQL at large $N$. In sharp contrast, the generation of metrologically useful ERSs rarely degrades even under substantial disorder in $\phi_{ij}$, consistently preserving the predicted HL scaling. Crucially, as detailed in Supplementary Materials \cite{supplement}, we analytically prove that this resilience would be a universal feature of the ERS generation, independent of the specific disorder distribution. Such inherent resilience—particularly in the large-system limit—positions the present framework as a promising candidate for architectures plagued by parameter inhomogeneity. Potential applications range from superconducting circuits with non-uniform couplings and calibration offsets \cite{Arute2019,Andersen2025} to disordered platforms like waveguide QED systems \cite{Zhang2025, Lodahl2015}, solid-state spins \cite{Davis2023,Lei2025} and ultracold polar molecules \cite{Gregory2024} where positional and interaction randomness are often intrinsic constraints.

\section*{Summary and outlook}
Our work fundamentally shifts the paradigm for accessing quantum advantage in metrology by uncovering physically significant manifolds hidden within the deep Hilbert space via engineered randomness. In particular, we demonstrate that Heisenberg-limited scaling can be ubiquitously unlocked in exponential-dimensional manifolds. It challenges the long-standing consensus that  many-body entangled states capable of offering quantum advantage are exceptional and restricted to the symmetric subspace of polynomial dimensionality. This finding establishes engineered randomness as a guiding principle for navigating the extremely challenging, exponentially complex Hilbert space, offering a versatile framework to probe its underlying geometry and structure while pushing the boundaries of both quantum information science and many-body physics.

Looking forward, our results open several compelling avenues extending far beyond metrology. The utility of our framework rests on the QFI, which serves dual roles as a metrological figure of merit and a rigorous witness of genuine multipartite entanglement \cite{Hauke2016}, a central yet elusive resource in quantum many-body physics \cite{Guhne2009}. Transcending a sole focus on the QFI, incorporating complementary metrics—such as the geometric measure of entanglement \cite{Gross2009,Bremner2009} or contextuality \cite{Howard2014}—might further illuminate additional physically significant regions of the Hilbert space. The framework of engineered randomness also suggests a new strategy for quantum circuit design: rather than aiming for Haar randomness \cite{Brandao2016,Schuster2025}, one could possibly tailor the measure over circuit parameter spaces to steer quantum evolutions toward valuable sectors of the Hilbert space.

\bibliography{scicite}
\bibliographystyle{scicite}

\section*{Acknowledgments}
This work was supported by the National Natural Science Foundation of China (12425414, U25D9006, 12174138, 12304572), Quantum Science and Technology-National Science and Technology Major Project (2024ZD0300900,2024ZD0300902), and the German Research Foundation (DFG) via project492547816 (TRR 360). Y.C. also acknowledges the support of the Fundamental Research Support Program of Huazhong University of Science and Technology (2025BRB001) and the Major Science and Technology Project of Hubei Province (2025BEA001). The authors declare no competing financial or non-financial interests.


\section*{Data Availability}
The data that support the findings of this study are included in this article and are available from the corresponding author upon reasonable request.

\newpage
\section*{Appendix}
The introduced $\alpha$-random unitary ensembles [see Eq.\,\eqref{Eq:First-Moment} in the main text] can generate metrologically useful states attaining Heisenberg-limited precision, rigorously supported by Theorem \ref{Theorem:1} and the concentration-of-measure phenomena in Theorem \ref{Theorem:2}  as well as the two corollaries for two explicit class of ERSs. Below, we present the key ingredients underlying our main results, showing how engineered randomness serves as a promising approach for exploring the exponentially vast Hilbert space.  We also detail the implementation of time-reversal quantum-enhanced metrology using engineered random states (ERSs) in a trapped-ion platform.

\subsubsection*{A. Proof of Theorem 1: average Heisenberg-limited scaling}
In Heisenberg picture, for an arbitrary operator $\mathcal{A}$, the characteristic first-moment structure of $\alpha$-random unitary ensembles in Eq.\,\eqref{Eq:First-Moment} yields Lemma 1,  
\begin{equation}
\label{Eq:OCprove}
\begin{aligned}
\mathbb{E}_{U\sim \mathcal{U}_\alpha} [U^\dagger \mathcal{A} U] &=  \sum_{i_1,i_2,j_1,j_2} \mathbb{E}_{U\sim \mathcal{U}_\alpha} [U_{i_1 j_1} U^{*}_{i_2,j_2}] \mathcal{A}_{i_2 i_1} |j_2\rangle\langle j_1|\\
& = \sum_{i_1,i_2,j_1,j_2} \alpha \delta_{i_1 j_1}\delta_{i_2 j_2 }\mathcal{A}_{i_2 i_1} |j_2\rangle\langle j_1|+(1-\alpha)\sum_{i_1 i_2 j_1 j_2} \Delta_{i_1,i_2,j_1,j_2} \mathcal{A}_{i_2 i_1} |j_2\rangle \langle j_1|\\
&= \alpha \mathcal{A} + (1-\alpha) \frac{\mathrm{tr}(\mathcal{A})}{D}\mathbb{I}.
\end{aligned}
\end{equation}
For convenience, we define a function over the high-dimensional space of $U$,
\begin{equation}
\label{Eq:fU}
f(U; \mathcal{A}) =\langle \Psi_U| \mathcal{A} |\Psi_U\rangle= \langle \Psi|U^\dagger \mathcal{A} U|\Psi\rangle,
\end{equation}
which evaluates the expectation value of $\mathcal{A}$ after the action of $U$ on the initial state $|\Psi\rangle$. Averaging over the ensemble $\mathcal{U}_\alpha$ gives the simple identity
\begin{equation}
\label{Eq:AveragefU}
\mathbb{E}_{U\sim \mathcal{U}_\alpha} [f(U; \mathcal{A})] = \alpha \langle \Psi| \mathcal{A} |\Psi\rangle + (1-\alpha)\frac{\mathrm{tr}(\mathcal{A})}{D}.
\end{equation}
This result allows the ensemble-averaged QFI in Eq.\,\eqref{Eq:QFI}, associated with a traceless collective Hermitian generator $\mathcal{O} = \sum_{m=1}^N \mathcal{O}_m$, to be written as
\begin{equation}
\begin{aligned}
\mathbb{E}_{U\sim \mathcal{U}_\alpha} [F_Q] &=  4 \mathbb{E}_{U\sim \mathcal{U}_\alpha} [f(U; \mathcal{O}^2)]-\mathbb{E}_{U\sim \mathcal{U}_\alpha} [f^2(U;\mathcal{O})]\\
& = \alpha \langle \Psi| \mathcal{O}^2 |\Psi\rangle - \mathbb{E}_{U\sim \mathcal{U}_\alpha} [f^2(U;\mathcal{O})] + (1-\alpha)\frac{\mathrm{tr}(\mathcal{O}^2)}{D}.
\end{aligned}
\end{equation}
In high-dimensional spaces, the function $f(U;\mathcal{A})$ is often governed by the concentration-of-measure phenomenon, ensuring that $f(U;\mathcal{A})$ remains tightly clustered around its mean value in Eq.\,\eqref{Eq:AveragefU}. By assuming that $\mathbb{E}_{U\sim \mathcal{U}_\alpha}[f^2(U;\mathcal{O})]-\mathbb{E}^2_{U\sim \mathcal{U}_\alpha}[f(U;\mathcal{O})]\leq \Theta(N)$, the second term simplifies to the square of the mean, giving rise to
\begin{equation}
\begin{aligned}
\mathbb{E}_{U\sim \mathcal{U}_\alpha} [F_Q] =\alpha \langle \Psi| \mathcal{O}^2 |\Psi\rangle - \alpha^2 \langle \Psi| \mathcal{O} |\Psi\rangle^2+(1-\alpha)\frac{\mathrm{tr}(\mathcal{O}^2)}{D}+\Theta(N).
\end{aligned}
\end{equation}
This compact form directly reproduces Theorem \ref{Theorem:1} where $\langle \Psi|\mathcal{O}^2|\Psi\rangle=\langle \Psi|\mathcal{O}|\Psi\rangle^2=\lambda^2 N^2$ and $\mathrm{tr}(\mathcal{O}^2)/D = \Theta(N)$ holds.

\subsubsection*{B. Proof of Theorem 2: concentration of superior metrological usefulness}
Starting from Eq.\,\eqref{Eq:EVconcentration} in Theorem \ref{Theorem:2}, we firstly evaluate the variance of the expectation value function $f(U,\mathcal{O})$ associated with the generator of parameter change. We define $y = f^2(U;\mathcal{O})-\mathbb{E}^2_{U\sim \mathcal{U}_\alpha}[f(U;\mathcal{O})]$. By using that $\mathbb{E}_{U\sim \mathcal{U}_\alpha}[f(U,\mathcal{O})]=\alpha \langle \Psi|\mathcal{O}|\Psi\rangle = \alpha \lambda N$, the probability of $|y|>\epsilon$ is upper bounded by
\begin{equation}
\begin{aligned}
\mathrm{Pr}\left(|y| \geq \epsilon\right) &\leq \mathrm{Pr}\left(|f(U;\mathcal{O})-\mathbb{E}_{U\sim \mathcal{U}_\alpha}[f(U;\mathcal{O})]| \geq \frac{\epsilon}{2 \mathbb{E}_{U\sim \mathcal{U}_\alpha}[f(U;\mathcal{O})]}\right) \\
& \leq 2 \exp\left(-\frac{c \epsilon^2}{4 \alpha^2 \lambda^2 N^2 L_\mathcal{O}^2}\right) = 2 \exp\left(-\Theta\left(\frac{D\epsilon^2}{\mathrm{poly}(N)}\right)\right).
\end{aligned}
\end{equation}
This inequality strictly implies that the variance of $f(U,\mathcal{O})$, namely $\mathbb{E}_{U\sim \mathcal{U}_\alpha}[y]$, is exponentially small in the Hilbert space dimension. Next, we characterize the statistical behavior of the QFI by defining the following quantity,
\begin{equation}
\bar{F}_Q \equiv 4 \mathbb{E}_{U\sim \mathcal{U}_\alpha} [f(U;\mathcal{O}^2)] - 4 \mathbb{E}_{U\sim \mathcal{U}_\alpha}^2 [f(U;\mathcal{O})]=4 \alpha (1-\alpha) \lambda^2 N^2 + \Theta(N).
\end{equation}
We examine the concentration of the QFI around $\bar{F}_Q$ by bounding the deviation probability,
\begin{equation}
\begin{aligned}
\mathrm{Pr}\Big(\left|F_Q - \bar{F}_Q\right| \geq \epsilon\Big) &= \mathrm{Pr}\left(\left|4 f(U;\mathcal{O}^2) -4 \mathbb{E} [f(U;\mathcal{O}^2)] - 4 f^2(U;\mathcal{O})+4 \mathbb{E}^2 [f(U;\mathcal{O})]\right|\geq \epsilon\right )\\
&\leq  \mathrm{Pr}\left(\left| f(U;\mathcal{O}^2) -\mathbb{E} [f(U;\mathcal{O}^2)] \right|+ \left|  f^2(U;\mathcal{O})- \mathbb{E}^2 [f(U;\mathcal{O})]\right|\geq \frac{\epsilon}{4}\right )\\
&\leq  \mathrm{Pr}\left(\left|f(U;\mathcal{O}^2) -\mathbb{E} [f(U;\mathcal{O}^2)] \right|\geq \frac{\epsilon}{8}\right )+\mathrm{Pr}\left(\left|f^2(U;\mathcal{O})- \mathbb{E}^2 [f(U;\mathcal{O})]\right|\geq \frac{\epsilon}{8}\right )\\
&\leq  2 \exp\left(-\frac{c \epsilon^2}{64 L_{\mathcal{O}^2}^2}\right) + 2 \exp\left(-\frac{c \epsilon^2}{256 \alpha^2 \lambda^2 N^2 L_\mathcal{O}^2}\right)\\
& \leq \exp\left(-\Theta\left(\frac{D\epsilon^2}{\mathrm{poly}(N)}\right)\right).
\end{aligned}
\end{equation}
By further making a replacement of $\epsilon \to \epsilon \bar{F}_Q$ in the above result and noticing that $\bar{F}_Q =\Theta(N^2)$, we find that  
\begin{equation}
\begin{aligned}
\mathrm{Pr}\Big(\left|F_Q - \bar{F}_Q\right| \geq \epsilon \bar{F}_Q \Big)  \leq \exp\left(-\Theta\left(\frac{D\epsilon^2 \bar{F}_Q^2}{\mathrm{poly}(N)}\right)\right) = \exp\left(-\Theta\left(\frac{D\epsilon^2}{\mathrm{poly}(N)}\right)\right),
\end{aligned}
\end{equation}
which directly yields Theorem \ref{Theorem:2} by taking $\epsilon$ to be a constant close to one.

\subsubsection*{C. RPS manifold: generation and concentration}
Building on the condition in Eq.\,\eqref{Eq:alpha}, the first moments of random diagonal unitaries exploited in the first construction of $\alpha$-random unitary ensemble reads
\begin{equation}
\label{Eq:FM_DRU}
\mathbb{E}[\Phi_{i_1 j_1} \Phi_{i_2 j_2}^*]  = 
\begin{cases}
\alpha \delta_{i_1 j_1}\delta_{i_2 j_2}, & \mathrm{if} \quad i_1\neq i_2;\\
1, & \mathrm{if} \quad i_1 = j_1=i_2=j_2.
\end{cases}
\end{equation}
In a $N$-qubit system, by following the same procedure as in Eq.\,\eqref{Eq:OCprove}, the operator contraction in Lemma 1 can be exactly derived for quantum operators $\mathcal{A}\in \{J_x, J_x^2\}$ under the special setting of $V=\mathbb{I}$ in Eq.\,\eqref{Eq:Construction}.
For an initial state fully polarized along the $x$ axis, the above equation allows us to exactly express the ensemble-averaged QFI as
\begin{equation}
\label{Eq:meanQFIexact}
 \mathbb{E}_{U\sim \mathcal{U}_\alpha} [F_Q] = \alpha (1-\alpha) N^2 + (1-\alpha)N,
\end{equation}
where the second term originates from the fact that $\mathrm{tr}(J_x^2)/D=N/4$.

Corollary \ref{theorem:rps}—that most RPSs are metrologically useful—follows from the concentration-of-measure phenomenon in high-dimensional Gaussian spaces. Physically, this can be viewed as a manifestation of the law of large numbers: typical behavior dominates almost all realizations. Starting from RPS in Eq.\,\eqref{Eq:RPS}, we introduce
\begin{equation}
\begin{aligned}
f(\vec{\varphi}; \mathcal{A})\equiv \langle \Psi (\varphi_1,\cdots,\varphi_D)|\mathcal{A} |\Psi (\varphi_1,\cdots,\varphi_D)\rangle = \frac{1}{D} \sum_{ab}  \langle a|\mathcal{A}|b\rangle e^{i(\varphi_b-\varphi_a)},
\end{aligned}
\end{equation}
where $\vec{\varphi} = [\varphi_1,\cdots,\varphi_D]\sim \mathcal{N}(0,\sigma^2 \mathbb{I}_D)$ is a Gaussian random vector with zero mean and covariance matrix $\sigma^2 \mathbb{I}_D$. We  show in the Supplementary Materials that $f(\vec{\varphi}; \mathcal{O}): \mathbb{R}^D \to \mathbb{R}$ is a Lipschitz continuous function with its Lipschitz constant bounded as
\begin{equation}
\begin{aligned}
L = \mathrm{sup}_{\vec{\varphi}\in \mathbb{R}^D}\left\| \nabla f (\vec{\varphi}; \mathcal{A}) \right\|_2 \leq  \frac{N^{\gamma}}{\sqrt{D}},
\end{aligned}
\end{equation} 
where $\gamma=1$ for $\mathcal{A}=J_x$ and $\gamma=2$ for $\mathcal{A}=J_x^2$, respectively. Based on the Gaussian concentration inequality, the probability that $f(\vec{\varphi}; \mathcal{A})$ deviates from its mean by more than $\epsilon$ is exponentially suppressed by the huge dimension of Hilbert space,
\begin{equation}
\mathrm{Pr}\Big(\Big|f(\vec{\varphi}; \mathcal{A})-\mathbb{E}_{\vec{\varphi}\sim \mathcal{N}(0,\sigma^2 \mathbb{I}_D)}[f(\vec{\varphi}; \mathcal{A})]\Big|\geq \epsilon \Big) \leq 2 \exp\left(-\frac{D\epsilon^2}{2 N^{2\gamma} \sigma^2}\right).
\end{equation}
This result shows that $f(\vec{\varphi};\mathcal{A})$ is overwhelmingly concentrated around its mean value, ensuring that almost all RPSs exhibit the typical metrological usefulness based on Theorem \ref{Theorem:2}. 
That is, the probability of encountering a RPS that cannot achieve Heisenberg-limited scaling is exponentially suppressed in the Hilbert space dimension,
\begin{equation}
\label{Eq:corollary2}
\mathrm{Pr}\Big(F_Q(\vec{\varphi};J_x)\Big) < \Theta(N^2) \Big) \leq \exp\left[-\Theta\left(\frac{D}{\sigma^2}\right)\right], 
\end{equation}
where $F_Q(\vec{\varphi};J_x)$ denotes the quantum Fisher information associated with the collective spin generator $J_x$.

\subsubsection*{D. CSS manifold: generation and concentration}
A key simplification arises from the invariance of Haar measure under both left and right multiplication. By leveraging this property, the second construction of the $\alpha$-random unitary ensemble takes the compact form
\begin{equation}
U = V \hat{\Phi} V^\dagger,
\end{equation}
where $V$ is a Haar-random unitary on $\mathrm{SU}(D)$ and $\hat{\Phi}$ denotes the eigenvalue matrix of $\Phi$. The first moments of $U$ can then be evaluated as
\begin{equation}
\begin{aligned}
\mathbb{E}[U_{i_1 j_1} U^{*}_{i_2,j_2}]
&= \sum_{mn} \mathbb{E}\left[ V_{i_1 m} V_{j_2 n} V_{j_1 m}^* V_{i_2 n}^*\right] \hat{\Phi}_{mm}\hat{\Phi}_{nn}^*\\
& = \frac{\alpha D^2-1}{D^2-1} \delta_{i_1 j_1}\delta_{i_2 j_2} + \frac{D}{D^2-1} (1-\alpha )\delta_{i_1 i_2}\delta_{j_1 j_2},\\
\end{aligned}
\end{equation}
with the ensemble parameter $\alpha = |\mathrm{tr}(\Phi)|^2/D$, which coincides with the normalized spectral form factor of $\Phi$. The second line follows directly from the Haar-average over four unitary matrix elements, a result that is by now standard in random matrix theory. Crucially, this compact expression is consistent with Eq.\,\eqref{Eq:First-Moment}, and it yields an identical exact operator contraction as in Lemma \ref{lem}.

Theorem \ref{Theorem:1} in the main text establishes that the optimal ensemble contraction factor, $\alpha = |\mathrm{tr}(\Phi)|^2/D$, takes the value $1/2$, leading directly to $\mathbb{E}(f_Q) = N/4+1/2$ for the QFI density of CSS. We now show how this condition can be realized in a generic $N$-particle quantum system. Consider each particle to be a spin-$J$ object with local dimension of $d = 2J + 1$, and take $\Phi = e^{-iH}$ as a local perturbation, for instance,
\begin{equation}
H = \chi S_z^{(1)}\otimes \mathbb{I}\otimes \mathbb{I}\cdots \otimes \mathbb{I},
\end{equation}
where $S_z = \mathrm{diag}[J,J-1,\cdots,-J]$. Under such an assumption, we obtain that
\begin{equation}
\alpha = \left(\frac{\sin\left[(J+\frac{1}{2})\chi\right]}{d \sin\left(\frac{\chi}{2}\right)}\right)^2.
\end{equation}
For qubit systems ($J=1/2$), this reduces to $\alpha = [\cos(\chi/2)]^2$, implying that $\alpha = 1/2$ is achieved at $\chi=\pi/2$. For qutrit ($J=1$), one finds $\alpha = [1+2\cos (\chi)]^2/9$, yielding $\alpha = 1/2$ at $\chi \approx 55.9^\circ$. The fixed unitary $\Phi$ may also represent a collective rotation. As an illustrative example, we focus on the $N$-qubit system with $\Phi = e^{-i\chi J_z}$. In this case, the ensemble parameter becomes
\begin{equation}
\alpha = \frac{1}{D}^2 \left |\sum_{m=0}^N C_N^m e^{i \left(m-\frac{N}{2}\right)\chi} \right|^2 = \left[\cos\left(\frac{\chi}{2}\right)\right]^{\frac{N}{2}}, 
\end{equation}
which reaches its optimal value $\alpha = 1/2$ at $\chi = \sqrt{2\log 2/N}\approx \sqrt{1.4/N}$.

To establish that most CSS generated by our second construction are metrologically useful, we exploit the concentration-of-measure technique \cite{Anderson2009}. In the same way, we consider the function returning the expectation value of the operator $\mathcal{A}$ with respect to $|\Psi_U\rangle$,
\begin{equation}
f(V;\mathcal{A}) = \langle \Psi_U |\mathcal{A}|\Psi_U\rangle  =  \langle \Psi |V \Phi^\dagger V^\dagger \mathcal{A} V \Phi V^\dagger|\Psi \rangle.
\end{equation}
The function $f(V;\mathcal{A}): \mathrm{SU}(D)\to \mathbb{R}$ is Lipschitz continuous on the special unitary group with its Lipschitz norm bounded by  $\|f\|_{\mathrm{Lip}}\leq 4 \|\mathcal{A}\|_{\infty}$ \cite{supplement}. By \emph{Lévy’s lemma}, for any $\epsilon>0$,
\begin{equation}
\mathrm{Pr}\left(\left|f(V;\mathcal{A})-\mathbb{E}_{\mathrm{Haar}} [f(V;\mathcal{A})]\right|\geq \epsilon \right)
\leq 2\exp\left(-\frac{D \epsilon^2}{64 \|\mathcal{A}\|_{\infty}^2}\right),
\end{equation}
This inequality shows that, as the dimension $D$ of the Hilbert space increases, the value of $f(V)$ for a Haar-random $V$ is exponentially concentrated around its mean. Based on Theorem \ref{Theorem:2}, the QFI of the CSS class associated with the collective generator $\mathcal{O} = \sum_{m=1}^N \mathcal{O}_m$ obeys
the following concentration inequality,
\begin{equation}
\mathrm{Pr}\left(F_Q(|\Psi_U\rangle; \mathcal{O}) <\Theta(N^2) \right) \leq \exp\left(-\Theta\left(D\right)\right),
\end{equation}
which immediately implies the Corollary \ref{Theorem:SCRS} in the main text. Detailed proofs are presented in the Supplementary Materials.

\subsubsection*{E. Experimental verification of ERS-enhanced metrology}
We performed our experiments on IonQ's Forte-1 quantum processor, which utilizes trapped ions with all-to-all qubit connectivity and high-fidelity operations, see Table S1 in the Supplementary Materials for detailed information about the device. To generate the ERSs, we implemented the unitary $U = V e^{iH} V^\dagger$ on the initial state $|0\rangle^{\otimes N}$, where $V^\dagger = \exp(-i \pi J_y/2)$ serves as a global $\pi/2$ pulse preparing the system in the $|+\rangle^{\otimes N}$ basis. The entangling Hamiltonian $H$ in Eq.\,\eqref{Eq:RHE} was realized through a sequence of pairwise controlled-phase gates, $\mathrm{CP}_{1j}\equiv \mathrm{CP}[4\phi_{1j}] = \mathrm{diag}[1,1,1,e^{4 i \phi_{1j}}]$ between the first qubit and qubits $j = 2$ to $N$. These unitaries were transpiled into the hardware-native GPI, GPI2, and ZZ gate set and executed for three distinct sets of random interaction weights, see Table S2 in the Supplementary Materials.

To characterize the metrological usefulness, we employed a time-reversal protocol where, following the ERS preparation, a collective phase $\vartheta$ was encoded via the rotation $e^{i\vartheta J_z}$. The accumulated phase information was then mapped onto the return probability by applying the inverse unitary $U^\dagger$, namely
\begin{equation}
P(\vartheta) =|\langle \boldsymbol{x}|e^{-iH} e^{-i\vartheta J_x} e^{iH}|\boldsymbol{x} \rangle|^2= |\langle \boldsymbol{0}|U^\dagger e^{-i\vartheta J_z} U|\boldsymbol{0} \rangle|^2.
\end{equation}
Data were acquired by scanning $\vartheta \in [0, 0.5]$ across 50 evenly spaced points, each averaged over 100 experimental shots. The resulting signal was fitted to a damped sinusoidal function 
\begin{equation}
    P(\vartheta) = A\, e^{-b\vartheta} \cos (c\vartheta + d) + B,
\end{equation}
from which we estimate the value of $\vartheta$ from the regions where the fitted curve has a large slope. The phase measurement uncertainty due to the quantum projection noise associated with measuring $P(\vartheta)$ can be evaluated as
\begin{equation}
\delta\vartheta^2 = \frac{P(\vartheta) [1-P(\vartheta)]}{|\partial_\vartheta P(\vartheta)|^2},
\end{equation}
which provides a direct lower bound on the QFI density of the experimentally generated ERSs via the quantum Cramér–Rao bound, i.e. $f_Q\geq \delta\vartheta^{-2}/N$, with the metrological gain over the SQL estimated in decibels as $20 \log_{10}(\delta\vartheta_{\mathrm{SQL}}/\delta\vartheta) = 20 \log_{10}(\delta\vartheta^{-1}N^{-1/2})$ \cite{Liu2022}.

 \foreach \x in {1,...,26} 
 {\clearpage 
 \includepdf[page=\x]{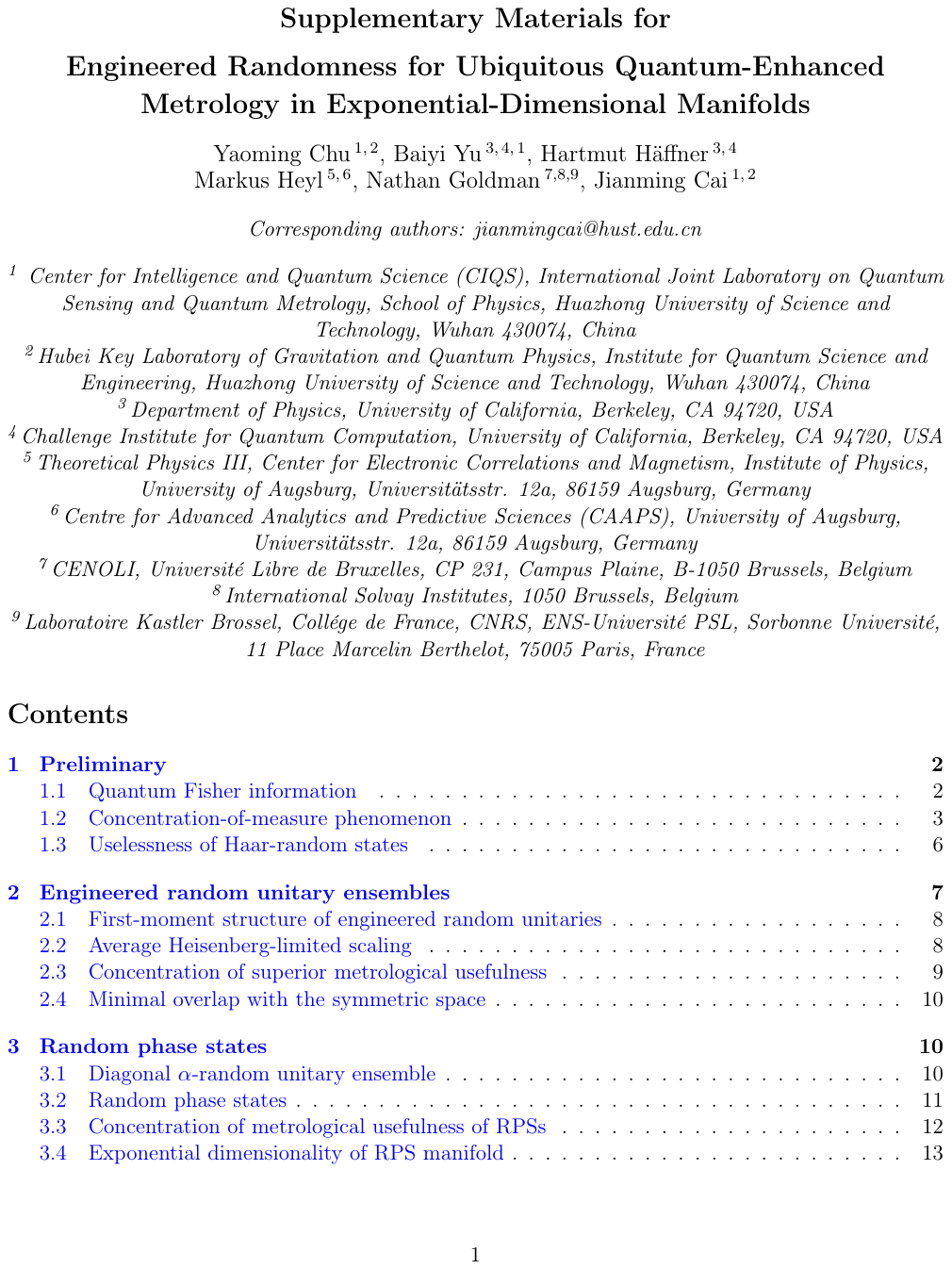}}

\end{document}